\documentclass[review]{elsarticle}
\usepackage{graphicx}
\usepackage[figuresright]{rotating}
\newcommand{\beq}[0] { \begin{eqnarray}}
\newcommand{\eeq}[0] { \end{eqnarray}}
\newcommand{\be}[0] { \begin{equation}}
\newcommand{\ee}[0] { \end{equation}}

\title{
Density Functional Theory Study of Solute Cluster Growth Processes 
in Mg-Y-Zn LPSO Alloys
}

\author[JAEA-k]{Mitsuhiro Itakura}
\author[JAEA-t]{Masatake Yamaguchi}
\author[UT]{Daisuke Egusa}
\author[UT,NIMS]{Eiji Abe}

\address[JAEA-k]{
Center for Computational Science \& e-Systems, Japan Atomic Energy Agency.
178-4-4 Wakashiba, Kashiwa, Chiba 277-0871, Japan}

\address[JAEA-t]{
Center for Computational Science \& e-Systems, Japan Atomic Energy Agency.
2-4 Shirakata-Shirane, Tokai-mura, Naka-gun, Ibaraki 319-1195, Japan}

\address[UT]{
Department of Materials Science and Engineering, 
University of Tokyo, Tokyo, Japan
}

\address[NIMS]{
Research Center for Structural
Materials, National Institute for Materials Science, Tsukuba, Japan}

\begin{document}

\begin{abstract}

Solute clusters in long period stacking order (LPSO) alloys play a key role
in their idiosyncratic plastic behavior, for example  kink formation and kink strengthening.
Identifying atomistic details of cluster structures is a prerequisite
for atomistic modeling of LPSO alloys and is crucial for
improving their strength and ductility; however,
there is much uncertainty
regarding interstitial atoms in the cluster. 
Although density functional theory calculations
have shown that the inclusion of Mg interstitial atoms is 
energetically most favorable in majority of LPSO alloys, 
solute elements have also been experimentally observed at interstitial sites.
To predict the distributions of interstitial atoms in the cluster
and to determine the kind of elements present,
it is necessary to identify mechanisms by which interstitial atoms are created.
In the present work, we  use density functional theory
 calculations to investigate growth processes of solute clusters, 
 specifically 
the Mg-Y-Zn LPSO alloy, in order to determine the precise atomistic structure of its solute clusters.
We show that a pair of an interstitial atom and a vacancy are spontaneously
created when a certain number of solute atoms are absorbed into the cluster,
and that all full-grown clusters should include interstitial atoms.
We also demonstrate that interstitial atoms are mostly Mg, while
the rest are Y; interstitial Zn atoms are negligible.
This knowledge greatly simplifies the atomistic modeling
of solute clusters in Mg-Y-Zn alloys.
Owing to the vacancies emitted from the cluster, vacancy density should be
super-saturated in regions where solute clusters are growing, and
increased vacancy density accelerates cluster growth.

\vspace{1pc}

\end{abstract}

\maketitle

\vspace{5pt}
\includegraphics[width=11cm]{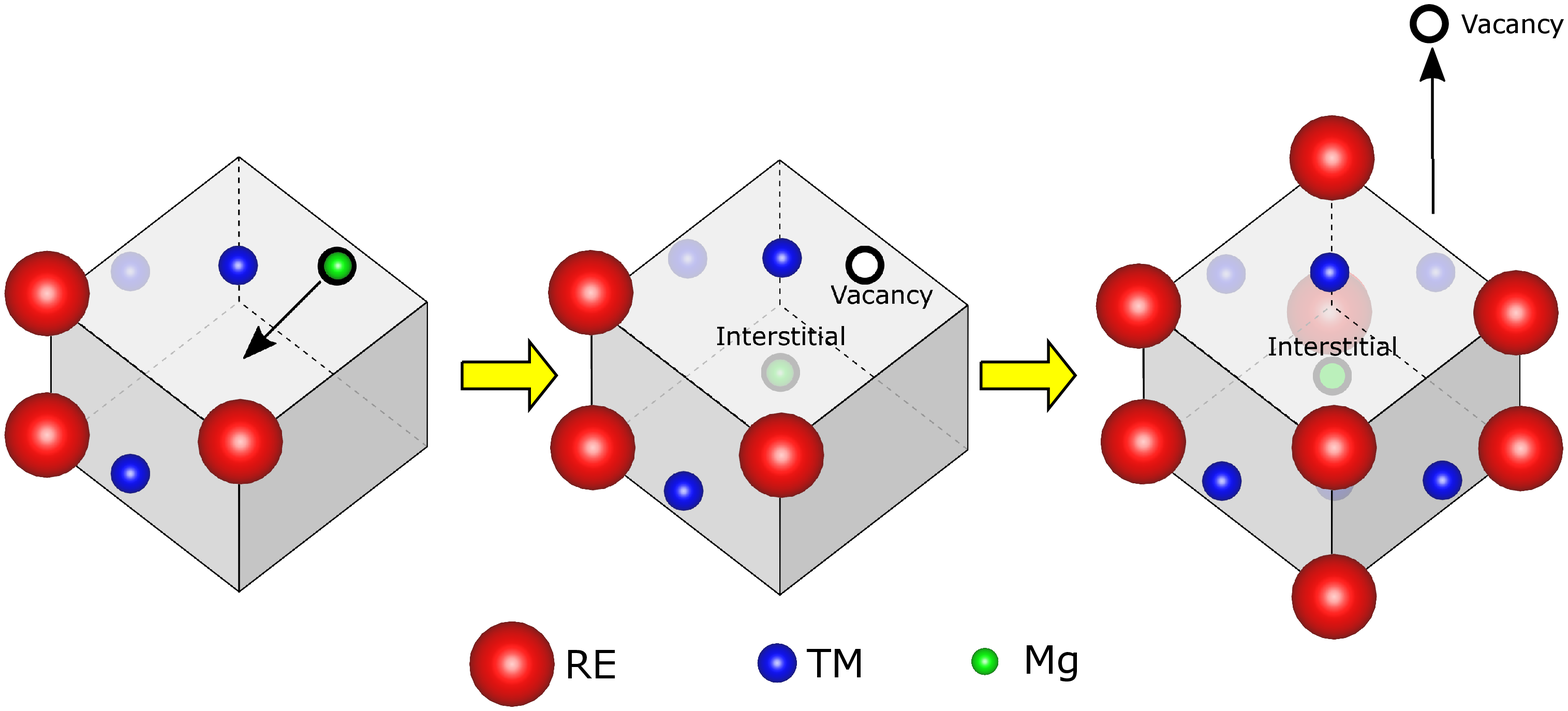}
\\
\centerline{\large Graphical Abstract}

\section{Introduction}

Solute clusters in Mg-based long period stacking order (LPSO)
alloys \cite{kawamura2001, abe2002} play a key role
in their idiosyncratic plastic behavior.
Solute atoms form $L1_2$-type clusters \cite{egusa2012} as shown in Fig. \ref{fig-L12}
and are strongly bound and displaced from their original lattice positions;
this makes it difficult for dislocations to cut through and allows only slips in the limited basal planes.
Although the lack of independent slip planes leads to poor ductility in most materials,
LPSO alloys exhibit so-called "kink deformation" \cite{hagihara2010,yamasaki2013,matsumoto2018},
which accommodates plastic deformation
in various directions {\it via} the structural organization of basal dislocations. 
It has been reported that these kink structures strengthen the material 
\cite{hagihara2010b,hagihara2013,xu2017,hagihara2019,somekawa2020}.
It is nonetheless expected that there is much room for improvement in terms of ductility and/or strength of
LPSO alloys {\it via} optimizing their compositions and the heat-treatment processes
through which they are formed.
Atomistic modelling of kink structures coupled with macroscopic models of plasticity
should give guiding principles for their improvement, and various models have been
proposed to account for kink strengthening
\cite{mayama2011,mayama2015,kobayashi2020,inamura2019}.

Scanning transmission electron microscopy (STEM)
 observations and density functional theory (DFT) calculations
have been used to reveal the atomistic structures of LPSO alloys,
including various stacking structures (10H, 12R, 14H, 18R, and 24R),
together with their thermodynamic stabilities
\cite{onorbe2012,kishida2012,saal2014, yamasaki2014,
    kishida2015,liu2015,wang2015,tane2015,kim2016,
    hagihara2016, liu2018,luo2020}
and the inter-cluster ordering along the stacking direction
 \cite{kishida2015,liu2018,luo2020,egusa2020}.
Nonetheless, there are uncertainties regarding the
interstitial atom (IA) in the cluster.
Although DFT calculations 
have shown that the inclusion of Mg-IA is 
energetically most stable in Mg-Y-Zn and Mg-Y-Ni LPSO alloys 
\cite{saal2014,liu2015,guo2020}, 
solute elements are also observed at interstitial sites
in Mg-Y-Zn alloy by STEM \cite{kishida2015}.
In contrast, for Mg-Y-Al LPSO alloy, DFT calculation indicates that Y-IA is energetically most
stable, and STEM observation have shown that about $80\%$ of interstitial 
sites are occupied by Y atoms \cite{egusa2020}.
To predict how much of the cluster has IAs
and which elements are present,
it is necessary to identify the mechanism by which IA is created,
because it is implausible to assume that a population of IAs is in
thermal equilibrium with its surroundings and can be estimated from
static formation energy alone. Rather, IA creation
should involve non-equilibrium processes.

Herein, we use DFT calculations to investigate 
 the growth processes of solute clusters, specifically 
that of the Mg-Y-Zn LPSO alloy, in order to identify the IA formation process.
The elementary process of cluster growth involves position exchange of solute atoms
mediated by vacancy migration. We calculated the energies of the initial and
final states of
various elementary steps and estimated the rate of occurrence of each process from 
these energies.
We also show that a pair of an IA and a vacancy are spontaneously
created when a certain number of solute atoms are absorbed into the cluster,
and that the full-grown cluster should include IAs.
Finally, we will show that IAs are mostly Mg atoms,
the rest being mostly Y atoms, while Zn IAs are negligible.

This paper is organized as follows. In Section 2, details of DFT calculations are described.
In Section 3, results of DFT calculations are presented. 
Section 4 discusses the consequences deduced from the obtained results.
In Section 5, a summary of the results and conclusions is presented.

\section{Details of the calculations}
\subsection{Atomistic model}

We investigated the formation energies of single solute clusters 
 embedded in a cell as shown in Fig.\ref{fig-cel}. 
Periodic boundary conditions were imposed in all directions, using a cell 
$4\langle \bar{1}010\rangle \times 4\langle 1\bar{1}00\rangle \times 5 \langle 0001\rangle$,
which contained 10 basal layers and 480 atoms.
For simplicity, we used the atomistic configuration of an almost
 isolated solute cluster, which corresponds to one quarter of the intra-plane cluster  density of a 
fully ordered LPSO structure, in order to avoid complexity arising from inter-cluster interactions. 

In the present work, we focus on the formation processes of solute clusters 
at the stacking fault (SF) region. Each cluster is separated from its periodic images
in the $\langle 0001 \rangle$ direction by 6 basal Mg layers without solute atoms.
We confirmed that
stacking order of
these Mg layers did not significantly affect cluster formation energy,
by comparing energies
calculated with two different stacking order in the margin area,
namely hexagonal 10H cell [ABABACBCBC][ABABACBCBC]$\cdots$
and rhombohedral cell [ABABACBCBC][BCBCBACACA]$\cdots$.
They differ by $50$ meV, which is only $1\%$ of total formation energy.
Although other choices for
the cell exist, such as 12R and 14H structures, 
the 10H structure is most convenient to preserve hexagonal 
symmetry in calculations.

Cell size was determined by minimizing the total energy of the 10H stacking structure
without any solute atoms. The calculated cell size was characterized by
lattice spacings  $a_0=3.19512$ and $c_0=5.18991$, which are slightly larger than those of hcp-Mg.
The cell size was fixed to these values for all calculations, allowing the total energy to be compared 
with a reference configuration.
Each case exerts different cell pressures, depending on the local expansion of the solute cluster configuration.
The maximum  cell pressure is approximately $40$ MPa. We estimated the correction 
to the total energy coming from the interaction between periodic images from cell pressure \cite{aneto},
finding it to be approximately $1$ meV. We elected to ignore this effect.

Discussions on the relationship between
SF formation and cluster growth have been reported,
since each phenomenon promotes the other and it is not clear
which takes place first \cite{okuda2015,mao2020}. In the case of Mg-Y-Zn,
direct observation of solute segregation near dislocation cores 
of Shockley partials and the subsequent growth of SF owing to Suzuki effect has been made \cite{hu2016}.
Therefore, we assumed that the Y/Zn solute cluster grew on an existing SF and calculated its 
cluster formation energy in the SF region of the 10H structure.

\subsection{DFT calculations}
Electronic structure calculations and structural relaxation by force minimization in DFT calculations
 were performed using the Vienna Ab-initio Simulation Package (VASP) \cite{vasp1,vasp2} with the projector augmented wave method and ultrasoft pseudopotentials. The exchange correlation energy was calculated using the generalized gradient approximation (GGA) with the Perdew-Burke-Ernzerhof function \cite{pbe}.  The Methfessel-Paxton smearing method with 0.2-eV width was used. The cutoff energy for the plane-wave basis set was 360 eV, and the convergence of cluster binding energy  with respect to increasing cutoff was confirmed. Structural relaxation terminated when the maximum force acting on the movable degrees of freedom became less than $10$ meV/$\AA$.

For the hexagonal supercell, k-points were placed on a Gamma-centered mesh in the XY-plane to preserve hexagonal symmetry; the Monkhorst-Pack k-point mesh was used in the Z-direction. 
The number of k-points was $2\times 2 \times 2$ in all cases.
 We confirmed that the convergence of cluster-formation energy 
with respect to the increasing k-point number was rapid,
with errors of approximately $5$ meV.

\section{Results}

Throughout the paper, we treat a vacancy as a type of solute element denoted by ``V", 
and thus use expressions such as ``site A is occupied by V".
The formation energy of a cluster C containing 
$l$-Y atoms, $m$-Zn atoms, and $n$-vacancies
with respect to a configuration in which each solute atom 
is isolated in the bulk region
is calculated as follows:
\be
E_f= E(\mbox{C})-E(0) -l[E(\mbox{Y}_1)-E(0)] 
-m[E(\mbox{Zn}_1)-E(0)] -n[E(\mbox{V}_1)-E(0)],
\ee
where 
$E(\mbox{C})$ is the total energy of cluster configuration C 
embedded in a calculation cell of the 10H structure, 
$E(0)$ is the total energy of the same cell without
any solute atoms, 
and $E(\mbox{X}_1)$ is the total energy of the configuration
in which one Mg atom in the bulk region is substituted by solute X.
When a cluster contains an IA, its formation process involves pair creation of
an IA and a vacancy, as will be shown later. Therefore, its reference state
contains no IA and one less vacancy. 
Accordingly, its formation energy is given by
\be
E_f= E(\mbox{C})-E(0) -l[E(\mbox{Y}_1)-E(0)] 
-m[E(\mbox{Zn}_1)-E(0)] -(n-1)[E(\mbox{V}_1)-E(0)].
\ee
This equation holds
regardless of the element species of IA, as well as when the cluster contains 
no vacancy.

The segregation energy of element X in the SF region, denoted by $E_{SF}(X)$, 
 is defined as the energy change in which a single solute atom X moves from bulk
 to SF. We found that 
$E_{SF}(Y)= -0.10$ eV, 
$E_{SF}(Zn)= -0.01$ eV, and $E_{SF}(V)= 0.00$ eV.
Note that the formation energy given by Eqs. (1) and (2) include the effect of SF segregation.
To concentrate on the binding energy between solute atoms, we define cluster binding energy $E_b$ as follows:
\be
E_b = E_f - \sum_X N_{SF}(X) E_{SF}(X),
\ee
where $N_{SF}(X)$ denotes the number of solute atoms X in the SF region.

Table 1 shows the two-body solute-solute interaction energy evaluated using $E_b$.
One can see that both the nearest neighbor Y-Zn and the next nearest neighbor Y-Y
interactions are attractive, whereas all other interactions between solute atoms (excluding V) are very small. This result is
consistent with previous work calculated using either the hcp or fcc lattice \cite{kimizuka2013}.
Table 2 shows the binding energies of various clusters containing up to 6 solute atoms
together with a measure of the many-body interaction $E_b -E_b^{(2)}$, where
$E_b^{(2)}$ denotes 
cluster expansion energy calculated using only two-body interactions
up to the next nearest neighbor as follows:
\be
E_b^{(2)} = \sum_{i,j} E_b(r_{ij};X_i,X_j),
\ee
where the summation runs through all solute pairs 
(excluding IA for simplicity), $r_{ij}$ is the relative position of
atoms $i$ and $j$, and $E_b(r_{ij};X_i,X_j)$ is 
the two-body solute-solute interaction energy between elements $X_i$ and $X_j$
as shown in Table 1. If $r_{ij}$ is neither the nearest neighbor nor next-nearest 
neighbor, it is set to zero.

Table 2 also shows
the number of tetrahedral sub-clusters  made of one Y atom and three Zn/V atoms.
One can see that such tetrahedra exhibit significant many-body interactions through
which they gain binding energy. A fully formed Y$_8$Zn$_6$ cluster has eight such tetrahedra
and it is expected that cluster growth is promoted by the formation of such tetrahedra.

The results shown in Table 2 indicate that vacancies are bound to the cluster as strongly as
solute atoms, but the relations between V absorption and solute absorption to the cluster
requires careful consideration. Each migration of solute atoms is mediated by
V diffusion. When a cluster absorbs a solute atom at a specific site,
that site is first occupied by a vacancy, after which that vacancy switches position with
a neighbor atom. If the neighbor atom is a solute atom,
it is absorbed to the cluster.

The rates of V absorption to a specific site and V emission from that site,
denoted by  $R(V^+)$ and  $R(V^-)$, respectively,
are given as follows:
\beq
R(V^+)&=& R_0 C_V,\\
R(V^-)&=& R_0 \exp(-|E_b^V|/k T),
\eeq
where $R_0$ is a V jump frequency in bulk, $C_V$ is equilibrium V density in the bulk,
$E_b^V$ is the binding energy of V to the cluster (negative if attractive),
 $k$ is Boltzmann's constant, and $T$ is temperature.
Under conditions of thermal equilibrium, $C_V=\exp(-E_f^V/k T)$, where $E_f^V = 0.88$ eV is 
vacancy-formation
energy in bulk as evaluated by DFT calculations. 
If the absolute value of
V binding energy $|E_b^V|$ is smaller than $E_f^V$,
 the site is occupied by V with probability 
$R(V^+)/R(V^-)= \exp((|E_b^V| - E_f^V)/kT)$ in thermal equilibrium conditions. 
In that case, the absorbed V quickly leaves the cluster,
and we can ignore clusters containing vacancies
when investigating growth processes.

We consider the rate of V emission followed by solute absorption, together with that of inverse process, 
denoted by  $R(V^-S^+)$ and  $R(S^-V^+)$, respectively.
Their ratio 
is given as follows:
\beq
R(V^-S^+)/R(S^-V^+)=\frac{ C_S \exp( |E_b^S|/kT)}{C_V \exp( |E_b^V|)/kT)},
\eeq
where $C_S$ denotes solute concentration in the bulk
and $E_b^S$ denotes solute binding energy (negative if attractive).
The rate of solute absorption is then given by $R(S^+)=R(V^-S^+) R(V^+)/R(V^-)$.
Solute absorption is possible if $R(S^+) > R(S^-V^+)$, which gives
\be
C_S \exp( |E_b^S|/kT) > 1.
\label{rate1}
\ee
For example, when $C_S=0.01$ and $T=500$ K, Eq. (8) gives $E_b^S < -0.2$ eV.

When the absolute value of V-binding energy is greater than V-formation energy,
the site is mostly occupied by V. Assuming that solute binding energy
is weaker than V-binding energy, the rate of solute absorption and that of the inverse process
are given as follows:
\beq
R(V^-S^+)&=&R_0 C_S \exp( (|E_b^S| - |E_b^V|)/kT),\\
R(S^-V^+)&=&R_0 C_V.
\eeq
Solute absorption is possible if $R(V^-S^+) > R(S^-V^+)$, which gives
\be
C_S \exp((|E_b^S| -|E_b^V| +E_f^V)/kT) > 1.
\ee
For the typical case of $C_S=0.01$ and $T=500$ K, this equation gives $|E_b^V|-|E_b^S| < 0.68$ eV.
If this condition is satisfied, the site is initially occupied by V and mostly remains so,
until the solute atom enters the neighbor site by diffusion and V swaps its position with
the solute atom.

As a cluster grows, Zn atoms in inner sites (a through f) tend to be displaced outwards,
whereas Y atoms in outer sites (A through H) tend to be displaced inwards.
The octahedral interstitial space at the center of the cluster becomes 
larger as the cluster grows and, after some threshold, the interstitial
atom can be accommodated.
Figure \ref{fig-neb} shows the migration energy profiles of atoms
moving from an inner site to the interstitial site
as evaluated by the nudged elastic band method \cite{NEB} with seven images.
When three inner sites are occupied by Zn atoms, the Mg atom in the inner site
can move to the interstitial site with a much lower energy barrier than
that of vacancy migration in bulk, thereby
creating a pair comprising an IA and a vacancy with some energy gain.
The Zn atom at the inner site can also move into the interstitial site, but
will be subject to a much larger energy barrier and less energy gain.

Table 3 shows the binding energies of various cluster configurations
and the energy gain of IA-V pair creation $\Delta E_{\mbox{IA}}$,
which gives the difference in $E_b$ between the cluster configurations
before and after IA-V pair creation.
In cases where one configuration is unstable and relaxes to
the other, 
we adopted a configuration in which the position of IA,
or the atom at an inner site that becomes IA, is fixed  
in structural relaxation, in order to evaluate the energy of an
unstable structure
by preventing the spontaneous creation or
annihilation of the IA-V pair during relaxation.
In addition, the position of the inner atom which is most distant from the IA
is also fixed to prevent the parallel transport of atoms.

As the cluster grows, the energy barrier becomes lower and the energy gain
becomes larger. When four or more inner sites are occupied by Zn atoms, 
the energy barrier becomes zero and configurations
without IA become unstable. From these results, we conclude that every
fully-grown cluster should contain IA.
The IA atom is most likely to be Mg; the Zn IA is very unlikely
because it has a much greater energy barrier for its creation compared to Mg.

Whereas Mg and Zn atoms have two stable positions in the inner site and interstitial site,
 we found that the Y atom occupying the inner site has only one stable position, which
moves toward the interstitial site as other inner sites are occupied by Zn atoms 
as shown in Fig. \ref{fig-yin}. Therefore, it is possible that when one inner site
is occupied by Y during the growth process, the Y atom becomes an IA. 
In Table 3, for the Y interstitial case, 
 the distance of the Y atom from the center, denoted by $r_Y$,
 normalized by the distance of an inner site from the center,
denoted by $r_0$, is shown. 
$\Delta E_{\mbox{IA}}$ for Y-IA case cannot be defined
because there is only one stable position for Y atom.
Precise estimation of the ratio between  Mg-IA and Y-IA requires 
detailed Monte Carlo simulations and is thus out of scope of the present paper.

Following the formation of an IA-V pair, V leaves the cluster if its binding energy
is not comparable to the V formation energy $E_f^V$, as discussed above.
If the binding energy of the newly created V is comparable to $E_f^V$,
it remains  until substituted by a solute atom, provided that
$C_S \exp((|E_b^S| -|E_b^V| +E_f^V)/kT) > 1$, as discussed previously.

Figure \ref{fig-ec} shows the progression of cluster-binding energy during the cluster growth process
forming the cluster Y$_8$Zn$_6$ plus IA.
For a given configuration, the energy gain of both Y and Zn absorption is investigated
at sites where large energy gains are expected {\it via} the formation of new attractive pairs
of solute atoms and/or a new solute tetrahedron composed of one Y and three Zn atoms.
Thereafter,
we chose a site and solute atom with maximum absorption energy, 
included that atom in the cluster, and repeated the process.
We confirmed that the solute binding energy satisfies 
either of the conditions $|E_b^V|<E_f^V$ or $C_S \exp((|E_b^S| -|E_b^V| +E_f^V)/kT) > 1$
in all cases.

Figure \ref{fig-de} shows
absorption energies at various sites, including ``incorrect" absorptions such
as Y absorption at inner sites and Zn absorption at sites other than 
inner or outer sites. Note that Zn absorption at outer sites does not create
new Y-Zn neighbor pairs and no energy gain is expected,  thus they were excluded
from  the calculation.
We evaluated only the case of Mg-IA.
One can see that, following the formation of IA, Y absorption at inner sites becomes
highly unfavorable, promoting the absorption of Zn atoms at inner sites.
Once the IA is created, ``correct" absorptions tend to increase the number of 
Y$_1$Zn$_3$ tetrahedra and are highly favorable.

\subsection{Effect of entropy on absorption}
Thus far, we have evaluated the absorption energies of solute atoms
to a cluster. Since cluster growth takes place at relatively high
temperatures, an entropy effect may modify absorption behavior.
In this subsection, we estimate the effect of entropy
from vibration and configuration.

Firstly, vibrational entropy is evaluated from the vibrational frequencies
of a single atom, either in bulk or in a fully formed
cluster with Mg interstitial atom.
Solute atoms included in a cluster are bound together by attractive
interactions and their amplitudes of vibration are expected to be smaller
than those in bulk, resulting in reduced entropy and increased free energy.
Vibration frequencies  $\omega_m$ 
of Mg, Y, and Zn atoms were
evaluated using DFT calculations and 
corresponding vibration free energies $F_{vib}$ 
as shown in Table \ref{tab:phonon}.
$F_{vib}$ was calculated using a quasi-harmonic approximation
\cite{qha} as follows:
\be
F_{vib} = \sum_{m=1}^3 \frac{\hbar \omega_m}{2} 
+ k T \ln(1-\exp \frac{-\hbar \omega_m}{k T}),
\ee
where $\hbar=1.0545718 10^{-34}$ (J$\cdot$s) 
denotes the reduced Planck constant.
One can see that the vibration frequencies of the Zn atom change
significantly
when it is absorbed into a cluster, whereas those of the Mg and Y atoms
remained 
virtually unchanged. However, its effect on free energy was at most
$0.08$ eV for the Zn atom, even at $600$K. Thus, we conclude that 
the effect of vibrational entropy on the absorption process is not significant.

Secondly, the configurational entropy of solute atoms is evaluated
by assuming that there a periodic array of solute clusters
and a uniform distribution of solute atom occur in bulk with concentration $C$.
Each cluster is surrounded by $N_b$ bulk Mg atoms, in which $N_b$ ranges from
$60$ to $100$ depending upon the periodicity of clusters.
When each cluster absorbs one solute atom, $C$ decreases by
$1/N_b$, and the
configurational entropy per atom, $S(C)= -k T [C\ln C +(1-C)\ln(1-C)]$,
decreases at $C<0.5$, and more rapidly for smaller values of $C$.
In the extreme case where $C$ becomes zero due to absorption and
entropy becomes zero, $C=1/N_b$ and
the change in entropy per cluster is given by
$ N_b S(1/N_b)$.
This term is positive and monotonically increases with increasing $N_b$.
For $T=500$K and $N_b=100$, the term
is $0.22$ eV. Thus, absorption energy
must be greater than $0.22$ eV at $T=500$. This condition is similar to 
Eq. (\ref{rate1}) 
derived from more simple considerations of absorption
and desorption rates.

\section{Discussion}
We have shown that 
a pair comprising an interstitial atom and a vacancy
is spontaneously created during the cluster growth process.
This entails that vacancy density becomes super-saturated in 
region where solute clusters are growing. 
In such a situation, it is possible that  vacancies accumulate
temporarily at the  solute cluster. A precise estimation of the effect 
of vacancy super-saturation on the cluster growth processes requires 
mesoscopic scale simulations of vacancy density evolution, and is thus outside of
the scope of the present work.

A more straightforward consequence is the acceleration of cluster growth processes,
since the elementary process of growth is vacancy diffusion and  the 
growth rate is directly proportional to vacancy density.
Normally, the growth rate of the LPSO structure is expected to be proportional
to $\exp(-(E_f^V+E_m^V)/kT)$ where $E_m^V$ is migration energy of vacancy.
In the super-saturated region, it can be accelerated to $\exp(-E_m^V/kT)$.

Another consequence of vacancy super-saturation is the promotion of
dislocation climb. When a Shockley partial dislocation absorbs 
vacancies, it gradually moves in the $\langle c \rangle$ direction.
Accelerated climb motion may assist in the formation of a periodic SF arrangement.

During the growth process of the LPSO structure, experimental observations
indicate that the period of stacking order becomes shorter,
and
solute clusters follow the migration of stacking faults
 \cite{onorbe2012}. 
If cluster migration occurs {\it via} the sequential migration of solute atoms,
each migration removes a solute atom from the cluster at a cost of approximately
$0.5$eV as shown in Fig. \ref{fig-de}, indicating that such migration is
quickly reverted.
It is more plausible to assume that a fully-grown cluster absorbs further
solute atoms to form a fused cluster \cite{kim2016}, which then emits extra
atoms and becomes an $L1_2$ cluster
again, but at different position.

\section{Conclusion}
Using DFT calculations, we have investigated the mechanisms of solute cluster growth
in Mg-Y-Zn LPSO alloys,  and have found that a tetrahedral cluster 
made of one Y atom and three Zn atoms is highly stable. A fully-grown cluster
contains eight such tetrahedra, and the formation of such tetrahedra is a strong driving force
for cluster growth. Similar calculations of other combinations of rare-earth and transition 
metal elements should reveal the origins of different LPSO structures with various compositions.
We also found that
a pair comprising an interstitial atom and a vacancy
is spontaneously created during this growth process, denoting that
every fully-grown cluster contains an interstitial atom.
The interstitial atom is most likely Mg, however, some portion can also be Y.
Interstitial Zn atoms should be negligible.
A vacancy created in the process is emitted from the cluster, and vacancy
density should thus become super-saturated during growth of the solute cluster.
This may promote dislocation climb and influence the evolution of the LPSO structure.
Mesoscopic modelling of cluster growth and vacancy emission, combined with
modelling of dislocation glide/climb and stacking fault growth, is expected
as the subject of future works.

\section{Acknowledgement}
The authors are grateful to Hajime Kimizuka for his useful comments.
This work was supported by JSPS KAKENHI for
Scientific Research on Innovative Areas "Materials Science
of a Mille-feuille Structure" (Grant Numbers 18H05480, 18H05479).
Computations were performed on the ICEX at the Japan Atomic Energy Agency.
The authors would like to thank Enago (www.enago.jp) for the English language review.

\clearpage

\newcommand{\myfig}[2] {
\begin{figure}[htb]
\vspace{9pt}
\includegraphics[width=14cm]{#1.eps}
\caption{#2} \label{#1}
\end{figure} }

\begin{table*}
\caption{
Two-body binding energy between solute atoms for nearest neighbor
and next nearest neighbor pairs in the stacking fault region.
Labels nn1, nn2, and nnn correspond to the  pairs shown in Fig.\ref{fig-cel}. 
Energies are given in meV.
}

\vskip 1em
\begin{tabular}{cccc}
Solute pair & 
$E_b$(nn1) &
$E_b$(nn2) &
 $E_b$(nnn) \\
\hline
 Y-Y &   
$+94$ &
$+151$ &
${\bf -80}$ \\
 Zn-Zn & 
$+10$ &
$+10$ &
$-13$ \\
 V-V &    
${\bf -96}$ &
${\bf -101}$ &
$-6$ \\
 Y-Zn &  
${\bf -52}$ &
${\bf -77}$ &
$ -6   $ \\
 Y-V &    
$+68$ &
$+35$ &
${\bf -41}$ \\
 Zn-V &
${\bf -35}$ &
${\bf -38}$ &
$+14$ \\
\end{tabular} \end{table*}

\begin{table*}
\caption{Binding energy $E_b$ for various clusters
consisting of up to six solute atoms. Each cluster is labelled according to 
the number of solute atoms (excluding vacancy) and its index number.
Letters A through H and a through f refer to the positions shown in
Fig. 1. 
$E_b^{(2)}$ is the binding energy estimated from cluster expansion
using pair interactions alone.
 All energies are in  meV. $N_T$ denotes  the
number of tetrahedra comprising one Y and three Zn or V.
The visualization of each configuration is given in the supplementary materials.
}
\vskip 1em
\begin{tabular}{lllcccc}
Label   &  Y & Zn & V & $E_b$ & $E_b - E_b^{(2)}$ & $N_T$ \\
 \hline

C3-1&G&ac&&$-213$ &$-56$ & \\
C3-2&b&ac&&$-202$ &$-49$ &\\
C3-3&B&ab&&$-174$ &$-45$ & \\
C3-4&B&ac&&$-132$ &$-29$ &\\
C3-5&BC&b&&$-207$ &$+1$ & \\
C3-6&BG&c&&$-234$ &$-4$ & \\

 \hline
C4-1&AB&ab &&$-377$ &$-41$ &\\
C4-2&ACH&b &&$-351$ &$+30$ &\\
C4-3&BG&ac &&$-406$ &$-46$ &\\
C4-4&G&ace &&$-487$ &$-250$ &1\\
C4-5&e &Edf&&$-445$ &$-214$ &1\\
C4-6&B&abc &&$-367$ &$-186$&1 \\
C3-7&B&ab&c &$-480$ &$-317$ &1\\

 \hline
C5-1&ABe&ab&&$-372$ &$+16$ &\\
C5-2&ABC&ab&&$-490$ &$-22$ &\\
C5-3&AB&abc&&$-574$ &$-185$&1 \\
C5-4&ACH&bd&&$-523$ &$-11$ &\\
C5-5&BG&abc&&$-647$ &$-210$ &1\\
C5-6&CH&bdf&&$-656$ &$-216$ &1\\

 \hline
C6-1&ABC&abc&&$-748$ &$-152$ &1\\
C6-2&ABCH&ab&&$-723$ &$+23$ &\\
C6-3&ABC&abe&&$-563$ &$-95$ &\\
C5-7&ABC&ab&d&$-569$ &$-101$ &\\
C5-8&ABC&ab&c&$-683$ &$-216$&1 \\
C6-4&ABCe&ab&&$-446$ &$+73$ &\\
C6-5&ABe&abc&&$-717$ &$-145$&1 \\
C6-6&BGd&ace&&$-864$ &$-171$&1 \\
C6-7&BGb&ace&&$-576$ &$-35$ &1\\
C6-8&BCG&abc&&$-813$ &$-167$ &1\\
C6-9&BG&abce&&$-939$ &$-674$ &2\\
\end{tabular} \end{table*}

\begin{table*}
\caption{Binding energy $E_b$ for various clusters with and without
an interstitial atom (IA). Each cluster is labelled according to
the number of solute atoms and its index number.
Letters A through H and a through f refer to the positions shown in Fig. 1.
A letter ``i'' in the label indicates that one atom
is located near an interstitial site.
$\Delta E_{\mbox{IA}}$ is the energy gain resulting from the creation
of the IA-V pair. All energies are in meV. 
Energies in parentheses indicate that the configuration is
unstable, and transforms
 to the structure designated by the labels shown together.
For the cases in which Y atom is located either at around an inner site
or an interstitial site,
the distance of Y from the octahedral interstitial site
is shown. See main text for details.
The visualization of each configuration is given in the supplementary materials.
}
\vskip 1em

\begin{tabular}{lcllclcc}
Label & IA & Y & Zn & V & $E_b$ & $\Delta E_{\mbox{IA}}$ \\
\hline
\hline
C4-1  &   &AB & ab&   &$-377$ & \\ 
C4-1i & Mg&AB & ab& d & $(+142)$ C4-1 & $+519$ \\ 
C5-3  &   &AB &abc&   &$-574$& \\ 
C5-3i & Mg&AB &abc& e & $-357$ & $+217$ \\ 
C6-1  &   &ABC&abc&   &$-748$&  \\
C6-1i & Mg&ABC&abc& e &$-808$ & $-60$ \\ 
C6-9  &   &BH&abce&   &$-939$ &  \\
C6-9i & Mg&BH&abce& d &$(-879)$ C6-9 &  $+60$ \\ 
C6-10 &   &AB&abcf&   &$-812$ & \\
C6-10i& Mg&AB&abcf& e &$-796$ & $+16$ \\ 
C7-1  &   &ABC&abcd&  &$-997$ & \\
C7-1i & Mg&ABC&abcd&e &$-1263$ & $-266$ \\
C8-1  &   &ABCD&abcd&  &$(-1168)$ C8-1i & \\ 
C8-1i & Mg&ABCD&abcd&e &$-1647$ & $-479$\\

\hline
C6-3  &   &ABC&abe&   &$-563$ & \\ 
C6-3i1& Zn&ABC&ab & e &$-547$ & $+16$ \\
C6-3i2& Mg&ABC&abe& f &$(-393)$ C6-3 &  $+170$ \\ 
C6-3i3& Mg&ABC&abe& c &$(-478)$ C6-3 &  $+85$ \\ 
C7-2  &  &ABC&abce&  & $-740$ & \\
C7-2i1&Zn&ABC&abc &e & $-980$ & $-240$ \\
C7-2i2&Mg&ABC&abce&d & $-1053$ & $-313$ \\
\hline
Label & IA & Y & Zn & V & $E_b$ &  $r_Y/r_0$ \\
\hline
C6-4&   &ABCe &ab&   & $-446$   &$0.90$ \\ 
C6-5&   &ABe  &abc&   &$-717$   &$0.84$ \\ 
C7-3&   &ABCe&abc &   &$-973$   &$0.73$ \\ 
C7-3i&Mg&ABCe&abc& d &$-419$    &$(\Delta E_{\mbox{IA}}=+554)$ & \\ 
C8-2i& Y &ABC&abcd& e &$-1411$  &$0.50$\\ 
C9-1i& Y &ABC&abcdf&e &$-2145$  &$0.37$ \\ 

\end{tabular} \end{table*}

\begin{table*}
\caption{Binding energy $E_b$ for various clusters with interstitial atoms (IA). 
Each cluster is labelled according to
the number of solute atoms and its index number.
Letters A through H and a through f refer to the positions shown in Fig. 1.
A letter ``i'' in the label indicates that one atom
is located at an interstitial site.
All energies are in meV. 
The visualization of each configuration is given in the supplementary materials.
}
\vskip 1em

\begin{tabular}{lcllccc}
Label & IA & Y & Zn & V & $E_b$ & $N_T$ \\
\hline
C6-1i & Mg & ABC    & abc & e & $-808$ & $1$ \\ 
C6-1i2& Mg & ABC    & abc & d & $-674$ & $1$ \\ 
C6-1i3& Mg & ABC    & abc &   & $-363$ & $1$ \\ 
C7-4i & Mg & ABC    & abce& &$-630$ & $1$ \\ 
\hline
C8-3i & Mg & ABCG   & abce& &$-1240$ &2\\
C8-4i & Mg & ABC    & abcde&&$-1026$ &1\\
C8-5i & Mg & ABCH   & abce& &$-1057$ &1\\
C8-6i & Mg & ABCD   & abce& &$-1048$ &1\\
\hline
C9-2i & Mg & ABCG   &abcde& &$-1667$ &3 \\
C9-3i & Mg & ABCGH  &abce & &$-1703$ &2 \\
C9-4i & Mg & ABCDG  &abce & &$-1710$ &2 \\
\hline
C10-1i& Mg & ABCGH  &abcde& &$-2296$ &3 \\
C10-2i& Mg & ABCDGH &abce & &$-2115$ &2 \\
\hline
C11-1i& Mg & ABCGH &abcdef& &$-3071$ &5 \\
C11-2i& Mg & ABCDGH&abcde & &$-2892$ &4 \\

C12-1i& Mg & ABCDGH&abcdef& &$-3680$ &6 \\
C13-1i& Mg & ABCDFGH&abcdef&&$-4257$ &7 \\
C14-1i& Mg & ABCDEFGH&abcdef&&$-4662$ & 8 \\ 
\hline
C9-1i &Y&ABC&abcdf&e& $-2145$& 3 \\ 
C9-1i1&Y&ABC&abcdf& & $-835$ & 3 \\
C10-3i&Y&ABC&abcdef&& $-1602$& 3 \\ 
C11-3i&Y&ABCD&abcdef&& $-2168$& 4 \\ 
C11-4i&Y&ABCG&abcdef&& $-2203$& 4 \\ 
C11-5i&Y&ABCH&abcdef&& $-2306$& 4 \\ 
C12-2i&Y&ABCDH&abcdef&& $-2825$& 5 \\ 
C12-3i&Y&ABCGH&abcdef&& $-2898$& 5 \\ 
C13-2i&Y&ABCDGH&abcdef&& $-3477$& 6 \\ 
C14-2i&Y&ABCDEGH&abcdef&& $-4029$& 7 \\ 
C15-1i&Y&ABCDEFGH&abcdef&& $-4623$& 8 \\ 
\end{tabular} \end{table*}

\begin{table}
\caption{
\label{tab:phonon}
Vibration frequencies of single atoms,
either in bulk or in fully formed clusters as
evaluated by DFT calculations, and
their contribution to free energy $F_{vib}$ at three temperatures
evaluated by quasi harmonic approximation.
Frequencies and free energies are in THz and meV, respectively.
}

\begin{tabular}{lcccrrr}
 &  & & & $F_{vib}$ & $F_{vib}$ & $F_{vib}$ \\
Atom & $\omega_1$ & $\omega_2$ &$\omega_3$ & $0$K & $300$K & $600$K \\
\hline
Mg bulk & 5.79 & 5.71 & 5.70 & 6 &-149 & -406 \\
Mg cluster & 5.37 & 5.36 & 5.32 & 5 & -155 & -417\\
\hline
Y bulk  & 4.01 & 3.92 & 3.90 & 4 & -178 & -464\\
Y cluster& 3.95 & 3.94 & 3.21 & 4 & -184 & -475\\
\hline
Zn bulk & 2.33 & 2.31 & 2.30 & 2 & -220 & -547\\
Zn cluster &4.18 & 4.12 & 3.58 & 4 & -178 & -464\\

\end{tabular}
\end{table}

\clearpage

\myfig{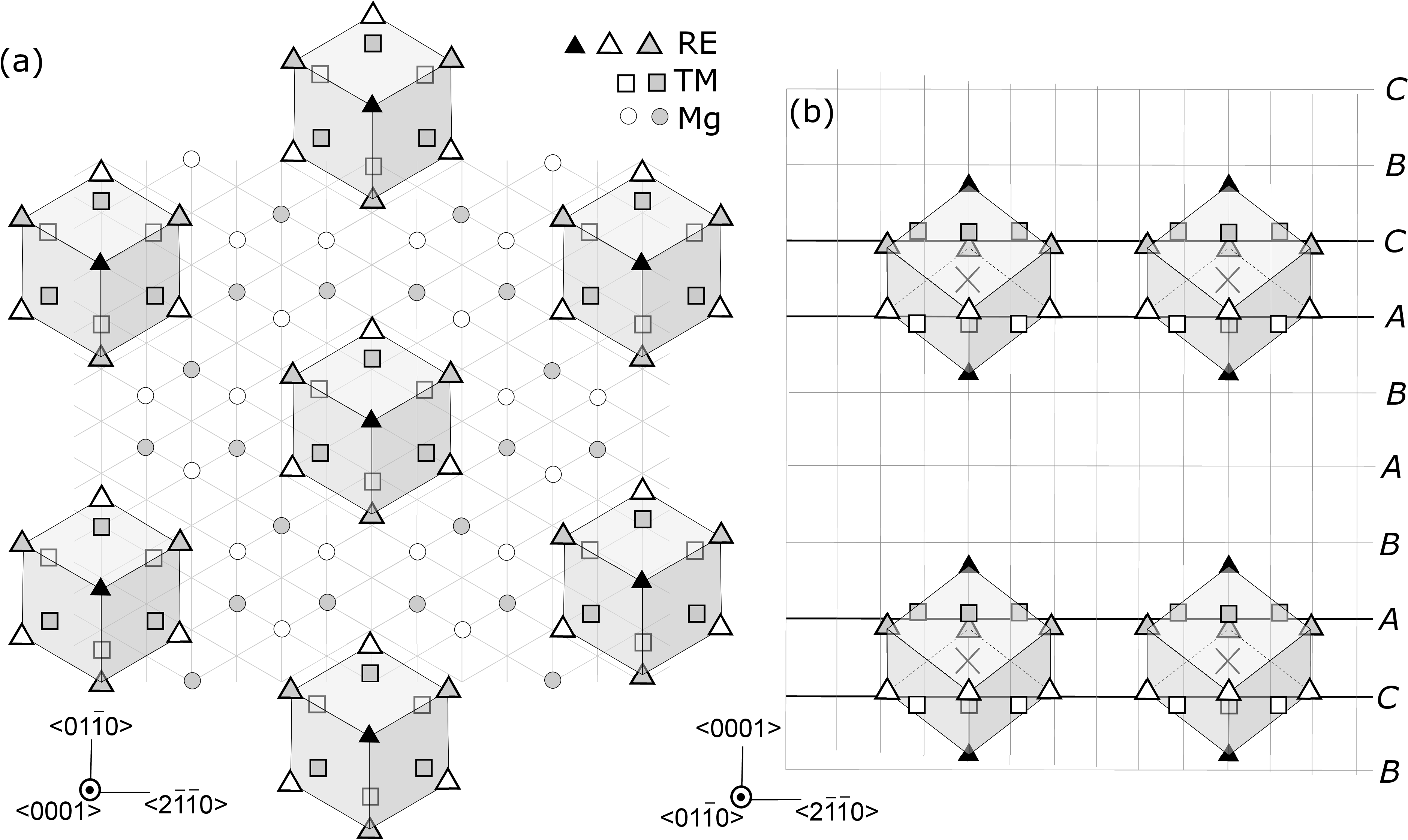} {
Typical atomistic structure of Mg-based LPSO alloys,
embedded in a 10H stacking structure, as seen from (a) $\langle 0001 \rangle$ and
(b) $\langle 01-10 \rangle$.
The circle, triangle, and square symbols  represent
Mg, rare-earth, and transition metal atoms, respectively.
The color of symbols (white, gray, and black) indicate different basal layers.
Each $L1_2$ cluster is embedded in a local fcc structure created by the
stacking faults shown as bold lines in (b). A, B, and C in (b) indicate
the order of stacking. Rare-earth atoms are located on the vertices of a cube
whereas 
transition metal atoms are located at the center of faces of a cube.
Rare-earth atoms are usually displaced toward the center of the cube
from the original lattice position, whereas transition metal atoms are
displaced away from the center.
}

\myfig{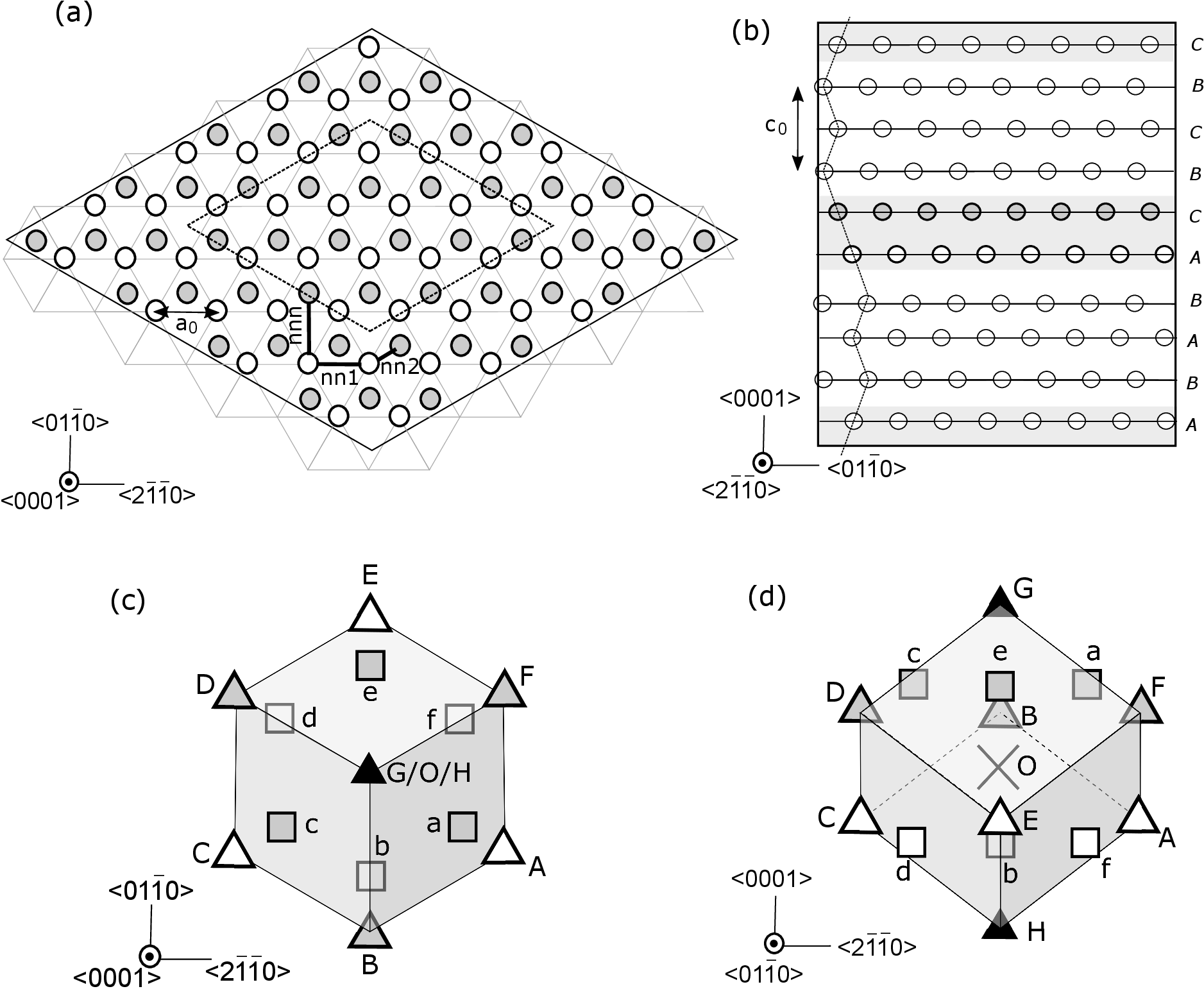} {
(a) and (b): Calculation cell used in the present work, consisting of 480 atoms.
Inner dashed line in (a) indicates the unit cell of the LPSO structure.
A, B, and C in (b) indicate the order of stacking and shaded areas
indicate stacking faults.
Pairs of atoms labeled as ``nn1'', ``nn2'', and ``nnn'' in (a) are pairs of
intra-plane nearest neighbor, inter-plane nearest neighbor, and
inter-plane next nearest neighbor, respectively.
The lattice spacing $a_0$ and $c_0$ is also shown in (a) and (b).
(c) and (d): atomistic configuration of the $L1_2$ cluster.
Letters ``A'' through ``H'' indicate sites occupied by Y atoms,
whereas letters ``a'' through ``f'' indicate sites occupied by Zn atoms.
The interstitial site is indicated by the letter O.
}

\myfig{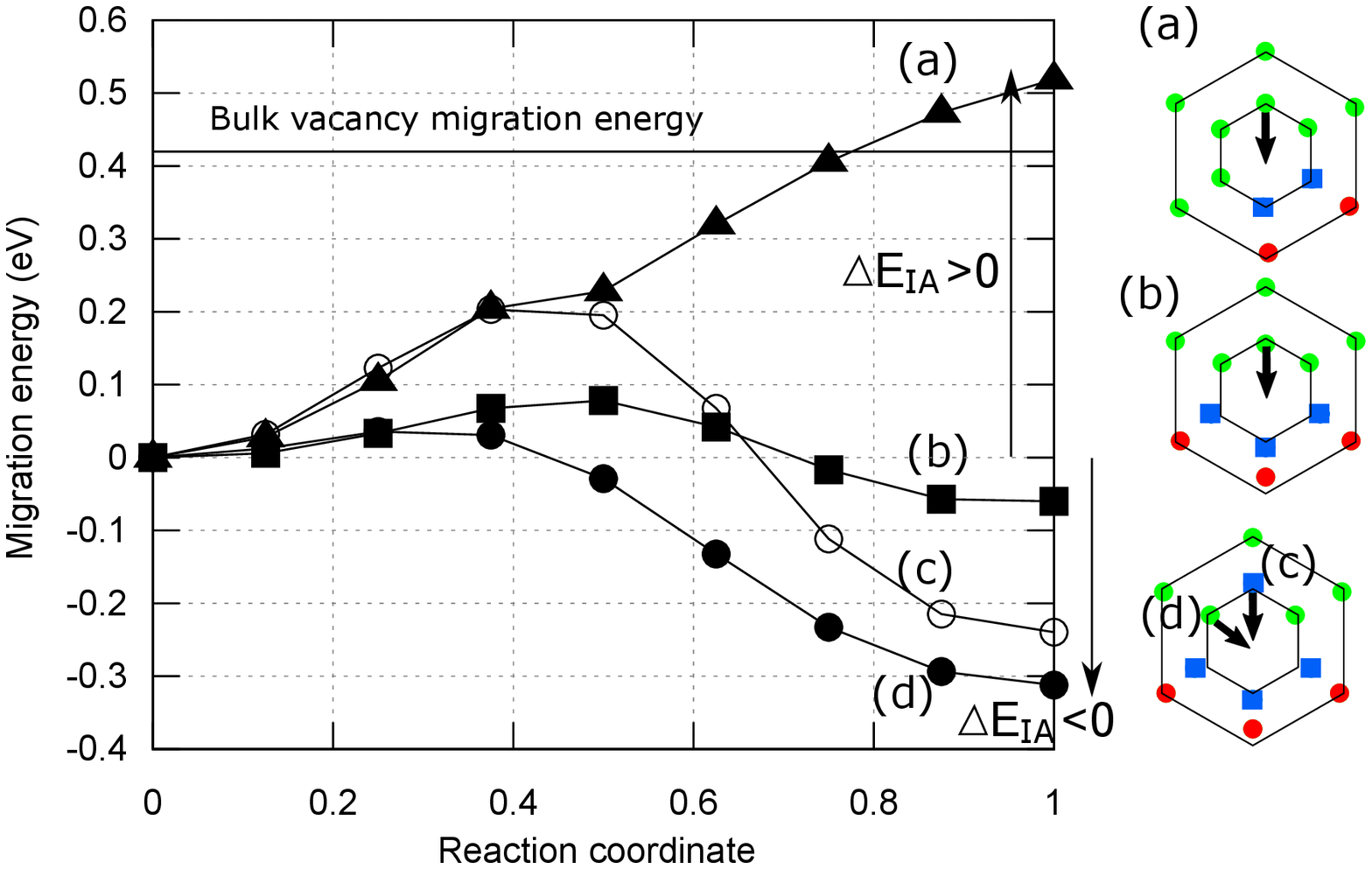} {
Energy profile of IA creation processes
evaluated by the nudged elastic band method.
The migration energy of a vacancy in hcp-Mg is also shown.
The atomistic configuration for each plot
(a) through (d) is shown at the right-hand side.
Green spheres, red spheres, and blue squares correspond to Mg, Y, and Zn atoms,
respectively.
Definition of interstitial atom creation energy $\Delta E_{\mbox{IA}}$ is schematically shown
by arrows.
}

\myfig{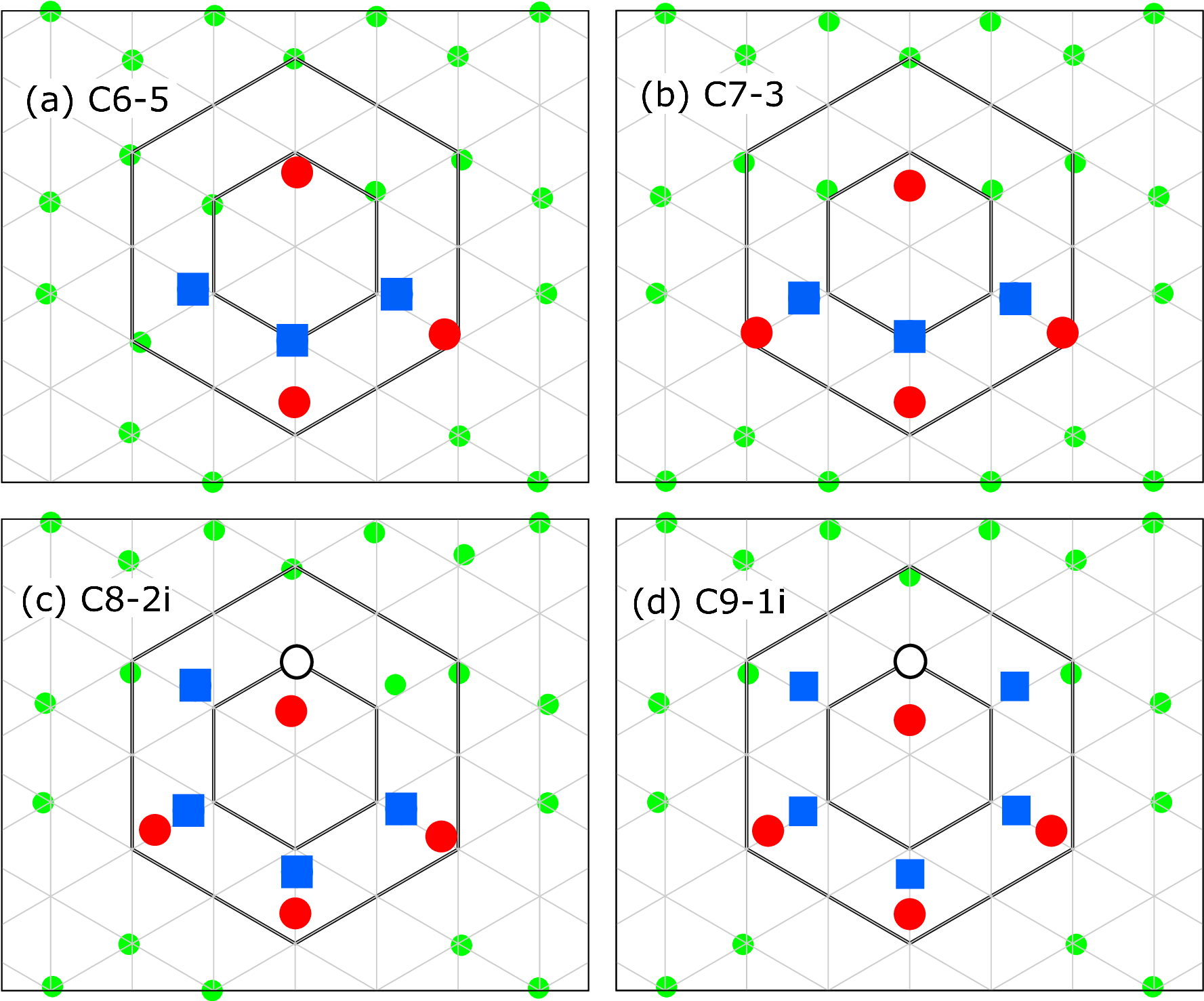} {
Atomistic configurations for several stages (a) through (d) of
cluster growth processes in which the Y atom gradually moves into
the interstitial position.
Green spheres, red spheres, and blue squares correspond to Mg, Y, and Zn atoms,
respectively; white spheres indicate vacancies.
}

\myfig{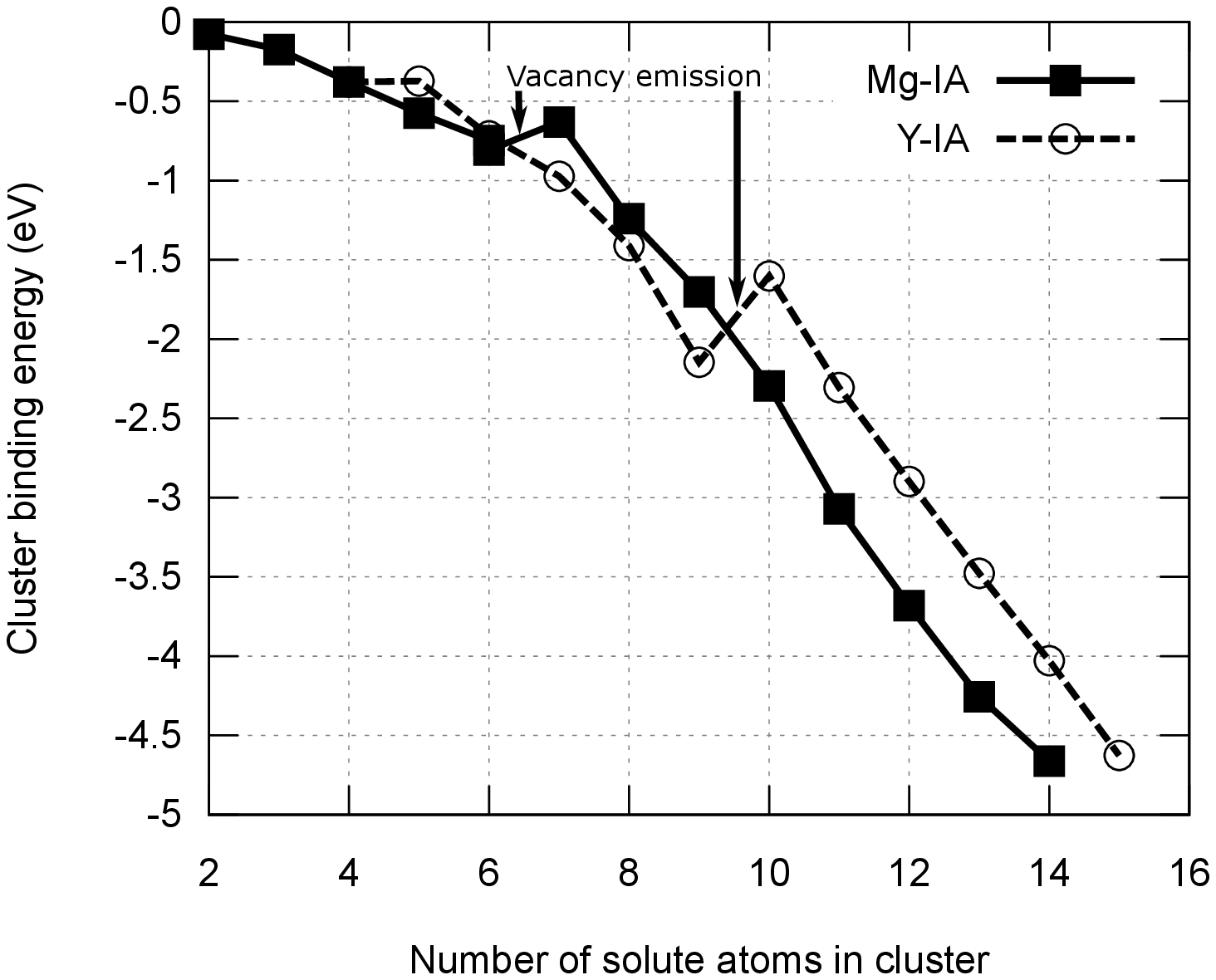} {
Cluster binding energy progression
during cluster growth processes
for two types (Mg and Y) of interstitial atom.
}

\myfig{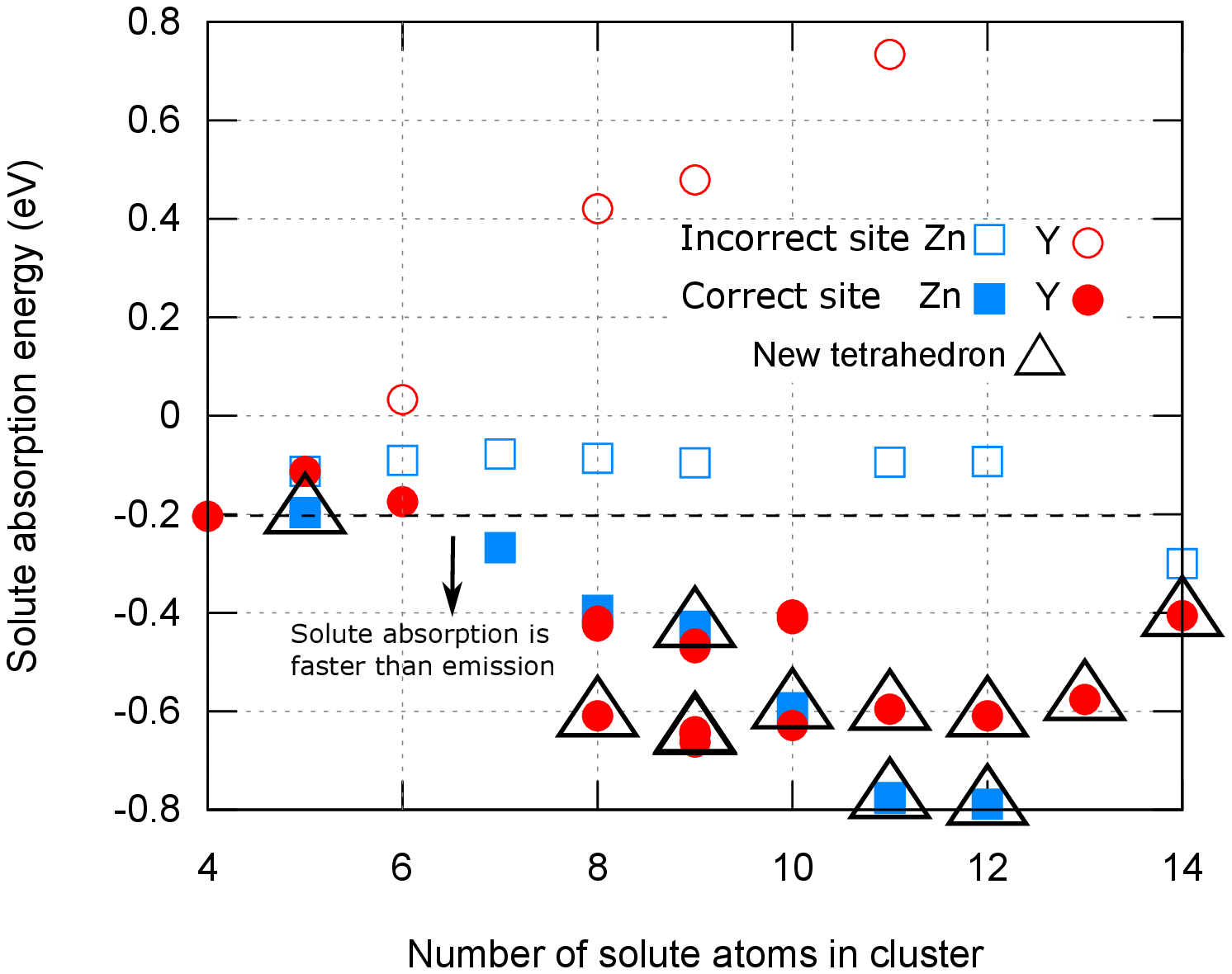} {
Solute absorption energy at various sites during the cluster growth.
"Correct" site absorption refers to the absorption of Zn atoms at inner sites
 and Y atoms at outer sites.
"Incorrect" site absorption refers to the absorption of Y atoms at inner sites
and Zn atoms at the nearest neighbor sites of outer sites, excluding the
inner sites.
Absorption processes causing an increase in the number of Y1Zn3 tetrahedra
are marked with triangles.
}

\end{document}